\documentclass{article}
\usepackage{graphicx} % Required for inserting images
\usepackage{amsmath}
\usepackage[affil-it]{authblk}
\usepackage{mathtools}
\usepackage{geometry}
\usepackage{float}
\newgeometry{top=3cm,left=3cm,right=3cm,bottom=3cm}
\usepackage{quantikz}
\usepackage{hyperref}
\usepackage{amssymb}
\usepackage{mhchem}
\usepackage{array}
\usepackage[super,sort,compress]{cite}
\usepackage{amsmath}
\usepackage{xcolor}
\usepackage{float}
\let\origfigure\figure
\let\endorigfigure\endfigure
\renewenvironment{figure}[1][H] {\origfigure[H]} {\endorigfigure}

\title{Improved cryptographic security in teleportation with q-deformed non-maximal entangled states  }

\author{P. Dasgupta \and D. Gangopadhyay }

\affil{ Department of Physics, Sister Nivedita University, DG Block , Newtown, Action Area 1, 
	
	Kolkata 700156, India }

\begin{document}

\maketitle

%\tableofcontents  % optional
\markboth{P.Dasgupta and D.Gangopadhyay}
{Teleportation with non-maximally entangled states }

\abstract{In this work the machinery of q-deformed algebras 
are used to enhance cryptographic security during teleportation. We use q-deformed harmonic oscillator states to develop a novel method of teleportation. The deformed states can be expressed in terms of standard oscillator states and the expressions contain certain arbitrary functions of $q$. It is the presence of these arbitrary functions that allows an enhancement of cryptographic security.
The specifics are :

(a) q-deformed Bell-like  states are constructed which reduce to the usual Bell states when the deformation parameter $q\rightarrow 1$. These deformed states form  an orthonormal basis for q-deformed entangled bipartite states when certain arbitrary functions of $q$ satisfy a constraint.

(b) We discuss the  generalisation of the usual teleportation protocol  with non-maximally entangled states. This generalisation  is then employed to construct two new protocols using   q-deformed non-maximally entangled  states. These states have additional parameters and these have to be shared for decryption after teleportation. Consequently, the cryptographic security is improved.}

\maketitle

 \section{Introduction:}
       
The concept of entanglement first materialized in the papers. \cite{EPR,Schro,Bell}. Quantum entanglement is associated with non-classical correlations between spatially separated quantum systems. Bell states form an orthonormal basis for an entangled bipartite qubit system:
	\begin{equation}
    \label{Bellstates}
		|\phi_0\rangle = \frac{|00\rangle+|11\rangle}{\sqrt{2}} ,
		|\phi_1\rangle=\frac{|01\rangle+|10\rangle}{\sqrt{2}} ,|\phi_{2}\rangle=\frac{|01\rangle-|10\rangle}{\sqrt{2}},
		|\phi_3\rangle=\frac{|00\rangle-|11\rangle}{\sqrt{2}}
	\end{equation}
	with $\langle\phi_i|\phi_j\rangle=\delta_{ij}$, $i,j=0,1,2,3$.

In this article we improve cryptographic security in teleportation using the novel formalism of q-deformed algebras \cite{Mac,Bied}. Using this formalism a harmonic oscillator realisation of deformed oscillators was first developed \cite{sir1,sir2,sir3}. Subsequently quantum logic gates were constructed with q-deformed oscillators \cite{qhad, qcnot}. It is to be noted 
that the references \cite{sir1,sir2,sir3,qhad,qcnot} were the first articles in the relevant fields. 

Recently new thumbrules for entanglement of bipartite qubit and qutrit systems were developed [\textit{P.Dasgupta, D. Gangopadhyay, arXiv 2505.21084 [quant-ph]}].
Here we use these constructions  for realisation of entangled states using q-deformed harmonic oscillators. Finally we show how a  teleportation protocol using non-maximal entangled states can be set up. This protocol has enhanced cryptographic security. Literature on various aspects of entangled states and non-maximally entangled states can be found in references \cite{15,16,17,18,19,20,21,22,23,24,25,26,27,28,29,30,31,32,33,34,35,36,37,38}  and references therein.

The plan of this article is as follows. Section 2 gives a brief review of q-deformed oscillators and the construction of q-deformed angular momentum states in the context of usual bipartite entangled states which has an underlying $SU(2)$ algebra. Section 3 deals with q-deformed bipartite states. In section 4 we set up analogues of Bell states using q-deformed oscillators. 
Section 5 describes our new protocol of teleportation using usual non-maximally entangled states. Section 6 describes this same new protocol of teleportation using \textit{non-maximally entangled q-deformed states.} Section 7 is the conclusion.  

\section{Review of q-deformed angular momentum states \cite{Mac,Bied,sir1,sir2,sir3,qhad,qcnot}}

A q-number \enskip[$x$]\enskip is defined as 
	\begin{equation}
    \label{q-num}
	   \enskip[x]\enskip=\frac{q^x-q^{-x}}{q-q^{-1}}
	\end{equation}
   Here $q=e^s$ , $0\leq s\leq 1$. For simplicity we will take $s$ to be real. In the limit $q\xrightarrow{}1$ , $\enskip[x]\enskip\rightarrow{} x$. That is, the usual number $x$ is recovered.

Let $a_q^{\dagger}$ and $a_q$ denote the  creation and annihilation operators for a q-deformed harmonic oscillator. The operators satisfy a deformed commutation relation (algebra)
\begin{equation}
\label{deformed_algebra_1}
a_qa_q^{\dagger}-qa_q^{\dagger}a_q=q^{-N_q}
\end{equation}
with 
\begin{equation}
\label{deformed_algebra_2}
  \enskip[N_q,a_q]\enskip=-a_q \enskip;\enskip
    \enskip[N_q,a_q^{\dagger}]\enskip=a_q^{\dagger} \enskip; \enskip
 a_q^{\dagger}a_q=N_q
  \end{equation}
 $N_q =N-(1/s)ln(\psi_2(q))I$ is the number operator of q-deformed harmonic oscillators, $N = a^{\dagger}a$ is the number operator for usual harmonic oscillators, $\psi_{2}(q)$ is an arbitrary function of $q$ which satisfies $\psi_{2}(q)=1$ for $q=1$ and $I$ is the identity operator.
   The q-deformed creation and annihilation operators $a_{q}^{\dagger} , a_{q}$ are 
	\begin{equation}
	a_{q}=a\sqrt{
		\frac{q^{N}\psi_1(q)-q^{-N}\psi_2(q)}{{N(q-q^{-1})}} }\enskip,\enskip	a_{q}^{\dagger}=\sqrt{
		\frac{q^{N}\psi_1(q)-q^{-N}\psi_2(q)}{{N(q-q^{-1})}} }a^{\dagger}
		\label{q-realization}
	\end{equation}
	$N$ is the number operator of the usual oscillator with eigenvalue $n$. The number operator corresponding to the q-deformed harmonic oscillator is $N_q$ and has eigen value $n_q=n-(1/s)ln(\psi_2(q))$. The functions $\psi_{1,2}(q)$ are arbitrary functions of $q$ and when $q=1$, $\psi_{1,2}(q)=1$. The presence of these arbitrary functions of $q$ provides us with an alternative formalism.  
	
	If $\psi_{2}(q)=1$ , then $N_q=N$ and deformed states cannot be distinguished from original states. However if that is not the case the occupation number becomes different for deformed and original oscillators.
	
	The Jordan-Schwinger construction gives a formalism of constructing generators of angular momentum states using two independent harmonic oscillators. 
If states are defined by total angular momentum $j$,  z-component $j_z$ and projection along $z$ by $m$ then
		\begin{equation}
        \label{usual jm}
		|jm\rangle=\frac{(a_1^{\dagger})^{j+m}(a_2^{\dagger})^{j-m}}{\enskip[(j+m)!(j-m)!]^{1/2}}|\Phi\rangle
	\end{equation}
Under q-deformation we get:

\begin{equation}
        \label{deformed jm}
		|jm\rangle_q=\frac{(a_{1q}^{\dagger})^{n_1}(a_{2q}^{\dagger})^{n_2}}{(\enskip[n_1]\enskip!\enskip[n_2]\enskip!)^{1/2}}|\Phi\rangle
	\end{equation}
    
Vacuum  state means absence of any particle. We take this vacuum state to be identical for both the usual oscillators and the q-deformed oscillators. This means that applying the usual creation operator on this vacuum will produce an usual oscillator state. Applying the q-deformed creation operator on this vacuum will produce a q-deformed oscillator state. Similar argument holds true for the destruction operators. $|\Phi\rangle=|\widetilde{0}\rangle=|\widetilde{0}\rangle_1|\widetilde{0}\rangle_2$ is the ground state ($j=0,m=0$), $j=(n_1+n_2)/2 $ ; and $ m=(n_1-n_2)/2$ where $n_{1,2}$ are eigenvalues of number operator of two usual oscillators and the $q-$ numbers \enskip[$n_{1,2}$]\enskip are the eigenvalues of the number operators of the two q-deformed harmonic oscillators.
	
For qubits, the only possible states corresponds to $(n_1+n_2)/2=1/2$  i.e $n_1=1-n_2$. Thus the states in occupation number representation will look like 
	
	$|(n_1+n_2)/2,(n_1-n_2)/2\rangle \equiv |n_1\rangle|n_2\rangle\delta_{n_1+n_2,1}$. Since $j=1/2$ for both qubits we suppress $j$ and so the states are
	
	\begin{equation} 
    \label{JS1}
		|m\rangle=\frac{(a_1^{\dagger})^{1/2+m}(a_2^{\dagger})^{1/2-m}}{\enskip[(1/2+m)!(1/2-m)!]^{1/2}}|\Phi\rangle \enskip; \enskip |-m\rangle=\frac{(a_1^{\dagger})^{1/2-m}(a_2^{\dagger})^{1/2+m}}{\enskip[(1/2+m)!(1/2-m)!]^{1/2}}|\Phi\rangle
	\end{equation}
	
        Equivalently in terms of $n_1,n_2$ we have,
	
	\begin{equation}
    \label{JS2}
	|n_1-1/2\rangle=\frac{(a_1^{\dagger})^{n_1}(a_2^{\dagger})^{1-n_1}}{\enskip[(n_1)!(1-n_1)!]^{1/2}}|\widetilde{0}\rangle \enskip;\enskip |-(n_1-1/2)\rangle=\frac{(a_1^{\dagger})^{1-n_1}(a_2^{\dagger})^{n_1}}{\enskip[(n_1)!(1-n_1)!]^{1/2}}|\widetilde{0}\rangle
	\end{equation}
Accordingly the qubits are:
\begin{equation}
\label{qubitoscillator}
\begin{split}
|0\rangle=|1/2,1/2\rangle=|1/2\rangle= a_1^{\dagger}|\widetilde{0}\rangle=a_1^{\dagger}|\widetilde{0}\rangle_1|\widetilde{0}\rangle_2=|\widetilde{1}\rangle_1|\widetilde{0}\rangle_2 \\ 
|1\rangle=|1/2,-1/2\rangle=|-1/2\rangle= a_2^{\dagger}|\widetilde{0}\rangle=a_2^{\dagger}|\widetilde{0}\rangle_1|\widetilde{0}\rangle_2=|\widetilde{0}\rangle_1|\widetilde{1}\rangle_2
\end{split}
\end{equation}

The column vectors denoting the usual computational basis states are
	$
	|0\rangle=\begin{pmatrix}
		1 & 0
	\end{pmatrix}^T,
	|1\rangle=\begin{pmatrix}
		0 &1
	\end{pmatrix}^T
	$  where  $\langle i|j \rangle=\delta_{ij}$ for (i , j=0 or 1).
We form one qubit with two independent oscillators where tilde states represent oscillator states and non tilde states represent qubits.
The ground state of harmonic oscillator is represented by $|\widetilde{0}\rangle$ whereas $|0\rangle$ , $|1\rangle$ represent qubits. These are the eigenvectors of $\sigma_3$.
 
Using equation (\ref{deformed jm}) and (\ref{JS2}) , the deformed qubits are given by :
\begin{equation}
\begin{split}
    | n_{1} -  1/2\rangle _{q}
\equiv {\frac {(a_{1q}^{\dagger})^{n_{1}} (a_{2q}^{\dagger})^{1-n_{1}}} 
 {(\enskip[n_{1}]\enskip!\enskip[1-n_{1}]\enskip!)^{1/2}}} | \Phi\rangle=(a_{1q}^{\dagger})^{n_{1}} (a_{2q}^{\dagger})^{1-n_{1}}|\Phi\rangle\\
| - (n_{1} -  1/2)\rangle_{q}
\equiv {\frac {(a_{1q}^{\dagger})^{1-n_{1}} (a_{2q}^{\dagger})^{n_{1}}}  {(\enskip[n_{1}]\enskip!\enskip[1-n_{1}]\enskip!)^{1/2}}}|\Phi\rangle=(a_{1q}^{\dagger})^{1-n_{1}} (a_{2q}^{\dagger})^{n_{1}}|\Phi\rangle
\end{split}
\end{equation}
Note that for $n=1,0$ we have $[n]=n$.

\section{The q-deformed bipartite state:}

A general bipartite state is
\begin{equation}
\label{usual_chi}
|\chi\rangle=\sum_{i,j=0}^1 a_{ij}|ij\rangle=a_{00}|00\rangle+a_{01}|01\rangle+a_{10}|10\rangle+a_{11}|11\rangle
\end{equation}
$a_{ij}$ are probability amplitudes with normalization $\sum_{i,j=0}^1a_{ij}^2=1$. It was shown in [\textit{arXiv 2505.21084 [quant-ph]}] that there exists a matrix $A=\begin{pmatrix}
    a_{00} & a_{01}\\
    a_{10} & a_{11}
\end{pmatrix}$ such that for $DetA \neq 0$ the state $|\chi\rangle$ is entangled while 
$DetA = 0$ means the state is unentangled.

$A$ matrices for each of the Bell States $|\phi_i\rangle$'s for $(i=0,1,2,3)$ are constructed by equating $|\phi_i\rangle=|\chi\rangle$ . For example, when
	 $|\chi\rangle= |\phi_0 \rangle$;
	$a_{00}=\langle 00|\phi_0 \rangle=1/\sqrt{2}$,
	$a_{01}=\langle 01|\phi_0 \rangle=0$, 
	$a_{10}=\langle 10|\phi_0 \rangle=0$, 
	$a_{11}=\langle 11|\phi_0 \rangle=1/\sqrt{2}$
	so that
	\begin{equation}
    \label{A0}
		A_0=\frac{1}{\sqrt{2}}\begin{pmatrix}
			1 && 0\\
			
			 0 &&1
\end{pmatrix}			 
		=\frac{I}{\sqrt{2}}
	\end{equation} 	
	Similarly, for $|\psi\rangle= |\phi_1 \rangle ,
    |\phi_2 \rangle, |\phi_3 \rangle$, the relevant matrices are respectively:
	\begin{equation}
    \label{A1-3}
		A_1
		=\frac{1}{\sqrt{2}}\sigma_1 ;\enskip
        A_2  
		=\frac{i}{\sqrt{2}}\sigma_2 ; \enskip
A_3
		=\frac{1}{\sqrt{2}}\sigma_3
\end{equation}
$I$ is the unit matrix and $\sigma_i$
	are Pauli Matrices.
	
  Commutation  and anticommutation relations satisfied by  these matrices are :
	\begin{equation}
	     \label{SU2-d}
        [A_i,A_j]=\sqrt{2} \sum_k (-1)^{i+j} \epsilon_{ijk}A_k \quad ; \{A_i,A_j\}=(-1)^{\frac{(i^3+j^3)-(i+j)}{4}}\delta_{ij}
    \end{equation}
 Under redefinition of variables :
	$A_1'=\sqrt{2}A_1$ , $A_2'=\sqrt{2}e^{(-i\pi)/2}A_2$, $A_3'=\sqrt{2}A_3$, 
	one obtains the $SU(2)$ algebra
	\begin{equation}
		\label{SU2}
		[A_i',A_j']=2i\epsilon_{ijk}A_k' \enskip; \enskip
        \{A_i ' , A_j ' \}=2\delta_{ij}
	\end{equation}
Consider a bipartite qubit state with  first qubit in Hilbert space $\mathcal{H}_A$ and second qubit in Hilbert space $\mathcal{H}_B$.The operators  $a_i$ and $b_j$ are usual annihilation operators in $\mathcal{H}_A$ and $\mathcal{H}_B$ respectively while  and $a_{iq}$ and $b_{jq} $ are their deformed counterparts. The  same holds for the  creation operators and  $i,j=1,2$. From equation \ref{q-realization} :    \begin{equation}
\label{a,b}
    \begin{split}
    a_{1q}^{\dagger}=\sqrt{\bigg(\frac{q^{N_1}\psi_1(q)-q^{-N_1}\psi_2(q)}{{N_1(q-q^{-1})}}\bigg)}a_1^{\dagger} \enskip;\enskip 
		a_{2q}^{\dagger}=\sqrt{\bigg(\frac{q^{N_2}\psi_3(q)-q^{-N_2}\psi_4(q)}{{N_2(q-q^{-1})}}\bigg)}a_2^{\dagger} \\ b_{1q}^{\dagger}=\sqrt{\bigg(\frac{q^{K_1}\beta_1(q)-q^{-K_1}\beta_2(q)}{{K_1(q-q^{-1})}}\bigg)}b_1^{\dagger} \enskip ; \enskip b_{2q}^{\dagger}=\sqrt{\bigg(\frac{q^{K_2}\beta_3(q)-q^{-K_2}\beta_4(q)}{{K_2(q-q^{-1})}}\bigg)}b_2^{\dagger} 
        \end{split}
			\end{equation}
			 with $ N_{1q}=N_1-(1/s)ln(\psi_2(q))$ , $ N_{2q}=N_2-(1/s)ln(\psi_4(q))$ , $ K_{1q}=K_1-(1/s)ln(\beta_2(q))$, 
			 $ K_{2q}=K_2-(1/s)ln(\beta_4(q))$. Here $N_i,K_j$ (for i,j=1,2) are number operators of the usual harmonic oscillators and $N_{iq} , K_{jq}$ are the deformed ones. As $\psi(q)$ and $\beta(q)$ are arbitrary we take
$\psi_1(q)=\psi_2(q)=\psi_3(q)=\psi_4(q)=\psi(q)$ and $\beta_1(q)=\beta_2(q)=\beta_3(q)=\beta_4(q)=\beta(q)$. So equation (\ref{a,b}) becomes 
            \begin{equation}
            \label{a's,b's}
            \begin{split}
a_{1q}^{\dagger}=\sqrt{\bigg(\frac{q^{N_1}\psi(q)-q^{-N_1}\psi(q)}{{N_1(q-q^{-1})}}\bigg)}a_1^{\dagger} \enskip ;\enskip
a_{2q}^{\dagger}=\sqrt{\bigg(\frac{q^{N_2}\psi(q)-q^{-N_2}\psi(q)}{{N_2(q-q^{-1})}}\bigg)}a_2^{\dagger}\\
b_{1q}^{\dagger}=\sqrt{\bigg(\frac{q^{K_2}\beta(q)-q^{-K_2}\beta(q)}{{K_2(q-q^{-1})}}\bigg)}b_1^{\dagger} \enskip;\enskip 
b_{2q}^{\dagger}= \sqrt{\bigg(\frac{q^{K_2}\beta(q)-q^{-K_2}\beta(q)}{{K_1(q-q^{-1})}}\bigg)}b_2^{\dagger}
                \end{split}
			\end{equation}
$ N_{1q}=N_1-(1/s)ln(\psi(q))$ ,
$ N_{2q}=N_2-(1/s)ln(\psi(q))$ ,
$ K_{1q}=K_1-(1/s)ln(\beta(q))$ 
and $ K_{2q}=K_2-(1/s)ln(\beta(q))$ . 

We now define the q-deformed bipartite state as a deformation of equation (\ref{usual_chi}) :
\begin{equation}
 |\chi\rangle_q=\sum_{i,j=0}^1\enskip[a_{ij}]\enskip|i\rangle_{Aq} \otimes |j\rangle_{Bq} =\enskip[a_{00}]\enskip|00\rangle_q+\enskip[a_{01}]\enskip|01\rangle_q+\enskip[a_{10}]\enskip|10\rangle_q+\enskip[a_{11}]\enskip|11\rangle_q
 \end{equation} 
 $|i\rangle_{Aq}$ is the first deformed qubit,
$|j\rangle_{Bq}$ is the second deformed qubit and $\enskip[a_{ij}]\enskip$ are the deformed amplitudes and equation (\ref{a's,b's} ) are the relevant creation and annihilation operator. (Note that $ \lim_{q \to 1}|\chi\rangle_q =|\chi\rangle$

Using equations (\ref{qubitoscillator}) and (\ref{a's,b's})  along with the fact that  the vacuum state is same for both usual and deformed oscillators we get:
 \begin{equation}
     \label{chi_q}
     \begin{split}
|\chi\rangle_q=\sqrt{\psi(q)\beta(q)}(\enskip[a_{00}]\enskip|00\rangle+\enskip[a_{01}]\enskip|01\rangle+\enskip[a_{10}]\enskip|10\rangle+\enskip[a_{11}]\enskip|11\rangle)\\=\sqrt{\psi(q)\beta(q)}\sum_{i,j=0}^1 \enskip[a_{ij}]\enskip|ij\rangle 
\end{split}
 \end{equation}
 $|ij\rangle$ denotes  the usual (not deformed)  bipartite qubit basis. That is the deformation effect has been transferred to the deformed amplitudes $[a_{ij}]$ and the arbitrary functions of $q$ , i.e. $\psi(q)$ and $\beta(q)$.

If the state $|\chi\rangle_q$ is normalized to $\enskip[M]\enskip$ {\it viz} ${}_q\langle \chi|\chi\rangle_q=\enskip[M]\enskip$, and $M$ belongs to the set of natural numbers, we get 

\enskip\enskip\enskip\enskip\enskip\enskip\enskip$\psi(q)\beta(q)=\frac{\enskip[M]\enskip}{\sum_{i,j=0}^1\enskip[a_{ij}]^2}$

Further 
\begin{equation}
\label{limit}
\begin{split}
\lim_{q \to 1
}|\chi\rangle_q =|\chi\rangle \\ =>
\lim_{q \to 1} \sqrt{\enskip[M]\enskip}\frac{\sum_{i,j=0}^1 \enskip[a_{ij}]\enskip|ij\rangle}{\sum_{i,j=0}^1 \enskip[a_{ij}]^2}=\sum_{i,j=0}^1
a_{ij}|ij\rangle\\=>
\sqrt{M}\frac{\sum_{i,j=0}^1 a_{ij}|ij\rangle}{\sum_{i,j=0}^1 a_{ij}^2}=\sum_{i,j=0}^1 
a_{ij}|ij\rangle \\ =>M=1 
\end{split}
\end{equation}
Note $\sum_{i,j=0}^1 a_{ij}^2=1$.
Therefore
\begin{equation}
    \label{psibeta}
    \psi(q)\beta(q)=\frac{\enskip[M]\enskip}{\sum_{i,j=0}^1\enskip[a_{ij}]^2} 
    =\frac{1}{\sum_{i,j=0}^1 \enskip[a_{ij}]^2}
\end{equation}
The explicit form of the deformed bipartite state is given in Equation (\ref{chi_q}) where $\{|00\rangle,|01\rangle,|10\rangle,|11\rangle\}$ are the usual (not deformed)  bipartite basis. We can now define the matrix [\textit{arXiv 2505.21084 [quant-ph]}].
\begin{equation}
\label{A_q}
A_q=\sqrt{\psi(q)\beta(q)}\begin{pmatrix}
    \enskip[a_{00}]\enskip&&\enskip[a_{01}]\enskip\\
\enskip[a_{10}]\enskip&&\enskip[a_{11}]\enskip
\end{pmatrix}
\end{equation}

For the state $|\chi\rangle_q$ to be unentangled $DetA_q=0$, otherwise the state is entangled for {\it all} non-zero values of $DetA_q $. For unentanglement (remembering that $\psi(q)\neq 0$ and $\beta(q)\neq 0$)
\begin{equation}
    \label{q-unentangle}
    \begin{split}
        DetA_q=0
        => \enskip[a_{00}]\enskip\enskip[a_{11}]\enskip=\enskip[a_{01}]\enskip\enskip[a_{10}]\enskip\\
        =>(q^{a_{00}}-q^{-a_{00}}).(q^{a_{11}}-q^{-a_{11}})=(q^{a_{01}}-q^{-a_{01}})(q^{a_{10}}-q^{-a_{10}})
    \end{split}
\end{equation}
There are a host of solutions to the above equation. Some of these are : $a_{00}=a_{01} \text{ and } a_{11}=a_{10}$ or $a_{00}=a_{10} \text{ and } a_{11}=a_{01}$ and so on.
\section{ The q-deformed Bell-like States:}
 We now define  q-deformed Bell-like states $|\phi_i\rangle_q$ as
 \begin{equation}
\label{q-bellstate}
    \begin{split}
|\phi_0\rangle_q=\sqrt{\psi(q)\beta(q)}\bigg([\frac{1}{\sqrt{2}}]\enskip|00\rangle + [\frac{1}{\sqrt{2}}]\enskip|11\rangle  \bigg) \\
|\phi_1\rangle_q=\sqrt{\psi(q)\beta(q)}\bigg(\enskip[\frac{1}{\sqrt{2}}]\enskip|01\rangle + [\frac{1}{\sqrt{2}}]\enskip|10\rangle\enskip\bigg) \\
       |\phi_2\rangle_q=\sqrt{\psi(q)\beta(q)}\bigg(\enskip\enskip[\frac{1}{\sqrt{2}}]\enskip\enskip|01\rangle - [\frac{1}{\sqrt{2}}]\enskip\enskip|10\rangle\enskip\bigg) \enskip\\
       |\phi_3\rangle_q=\sqrt{\psi(q)\beta(q)}\bigg(\enskip\enskip[\frac{1}{\sqrt{2}}]\enskip\enskip|00\rangle - [\frac{1}{\sqrt{2}}]\enskip\enskip|11\rangle\enskip\bigg)    \end{split}
\end{equation}\\
Note that that $\lim_{q \to 1}|\phi_i\rangle_q =|\phi_i\rangle$, $(i=0,1,2,3)$. $|\phi_i\rangle$ is defined in equation (\ref{Bellstates}) and when $q\rightarrow 1$, $\enskip[\frac{1}{\sqrt{2}}]\enskip\to \frac{1}{\sqrt{2}}$ and $\psi(q)\beta(q)\to 1$.
 Moreover equating $|\chi\rangle_q=|\phi_i\rangle_q$ and using equation (\ref{psibeta}) we get the orthonormality conditions for q-deformed Bell-like states. For example, if $|\chi\rangle_q=|\phi_0\rangle_q$, then, $\enskip[a_{00}]\enskip=\enskip[a_{11}]\enskip=\enskip[\frac{1}{\sqrt{2}}]\enskip$ and $\enskip[a_{01}]\enskip=\enskip[a_{10}]\enskip=\enskip[0]\enskip=0$ (0 and 1 have no q-deformation) and we get :
  \begin{equation}
 \label{q-bell-orthonorm}
    \psi(q)\beta(q)=\frac{1}{ 2\enskip[\frac{1}{\sqrt{2}}]^2}
 \end{equation}
 Equation (\ref{q-bell-orthonorm}) also holds when $|\chi\rangle_q=|\phi_i\rangle_q$ for $(i=1,2,3)$.

Consider a q-deformed entangled state 

$|\mu\rangle_q=\enskip[a_{00}]\enskip|00\rangle_q+\enskip[a_{11}]\enskip|11\rangle_q = \sqrt{\psi(q)\beta(q)}(\enskip[a_{00}]\enskip|00\rangle+\enskip[a_{11}]\enskip|11\rangle)$ such that $[a_{00} ]\neq 0$ and $[a_{11}] \neq 0$. 

$|\mu\rangle_q$ can be expressed as a linear 
superposition of q-deformed Bell-like states:
 \begin{equation}
 \label{lieanrsup}
     \begin{split}
|\mu\rangle_q =b_0|\phi_0\rangle_q+b_1|\phi_1\rangle_q+b_2|\phi_2\rangle_q+b_3|\phi_3\rangle_q\\
with\enskip\enskip  b_0=\frac{[a_{00}]+[a_{11}]}{2[1/\sqrt{2}]}\enskip,\enskip b_1=b_2=0 \enskip, \enskip b_3=\frac{[a_{00}]-[a_{11}]}{2[1/\sqrt{2}]}
     \end{split}
 \end{equation}  
Thus any q-deformed bipartite entangled state can be written as a linear superposition of q-deformed Bell-like states. The amplitudes associated with the q-deformed Bell-like basis are functions of q.

We now construct the $A_q$ matrices defined in equation
(\ref{A_q}) for q-deformed Bell-like states using 
equation (\ref{q-bellstate}).
If $|\chi\rangle_q=|\phi_0\rangle_q$, then $\enskip[a_{00}]\enskip=\enskip[a_{11}]\enskip=\enskip[\frac{1}{\sqrt{2}}]\enskip$, and  we get $A_{0q}$. 
Similarly when $|\chi\rangle_q=|\phi_i\rangle_q$ , we get the corresponding $A_{iq}$ matrices for $(i=1,2,3)$.
\begin{equation}
    \label{Aqs}
    \begin{split}
A_{0q}=\sqrt{\psi(q)\beta(q)}\begin{pmatrix}
    \enskip[\frac{1}{\sqrt{2}}]\enskip && 0\\
    0 && \enskip[\frac{1}{\sqrt{2}}]\enskip
\end{pmatrix}=\sqrt{\psi(q)\beta(q)}\enskip[\frac{1}{\sqrt{2}}]\enskip I
\\A_{1q}=\sqrt{\psi(q)\beta(q)}\enskip[\frac{1}{\sqrt{2}}]\enskip \sigma_1 \\A_{2q}=\sqrt{\psi(q)\beta(q)}\enskip[\frac{1}{\sqrt{2}}]\enskip i \sigma_2 \\A_{3q}=\sqrt{\psi(q)\beta(q)}\enskip[\frac{1}{\sqrt{2}}]\enskip \sigma_3.
   \end{split}
\end{equation}
$I$ is the 2-dimensional identity matrix and  
$\sigma_i,(i=1,2,3)$ are the Pauli matrices.
The commutation and anticommutation relations satisfied by the $A_{iq}$ (i=1,2,3) matrices are  given by:
\begin{equation}
    \label{qdefcom}
    \begin{split}
    \enskip[A_{iq},A_{jq}]\enskip=(-1)^{i+j} \sqrt{\psi(q)\beta(q)}\enskip \enskip[\frac{1}{\sqrt{2}}]\enskip\enskip2 \epsilon_{ijk} A_{kq}\\
     \{A_{iq},A_{jq}\}=(-1)^{\frac{(i^3+j^3)-(i+j)}{4}} \psi(q)\beta(q)\enskip[\frac{1}{\sqrt{2}}]\enskip2\delta_{ij}
    \end{split}
\end{equation}
 For the limiting condition $q\to 1$ , the above commutation ad anticommutation relations reduces down to $\enskip[A_{i},A_{j}]\enskip=(-1)^{i+j}\sqrt{2} \epsilon_{ijk} A_{k}$ and $ \{A_{iq},A_{jq}\}=(-1)^{\frac{(i^3+j^3)-(i+j)}{4}}\delta_{ij}$ respectively which has been already given in equation (\ref{SU2-d}).

 Redefining the $A_{iq}$ matrices as :
 \begin{equation}
 \begin{split}
     A_{1q}'=\frac{A_{1q}}{\enskip[\frac{1}{\sqrt{2}}]\enskip\sqrt{\psi(q)\beta(q)}}\\A_{2q}'=\frac{e^{-i\pi/2}A_{2q}}{\enskip[\frac{1}{\sqrt{2}}]\enskip\sqrt{\psi(q)\beta(q)}} \\
 A_{3q}'=\frac{A_{3q}}{\enskip[\frac{1}{\sqrt{2}}]\enskip\sqrt{\psi(q)\beta(q)}}
 \end{split}
 \end{equation}
  we get,  $\enskip[A_{iq}',A_{jq}']\enskip=2i\epsilon_{ijk}A_{kq}'$ which is already given in equation (\ref{SU2}) , which is the usual SU(2) algebra.
\section{Teleportation using usual undeformed non-maximally entangled states [\textit{arXiv 2505.21084 [quant-ph]}]}
\begin{figure}
\centering
\begin{quantikz} 
    \lstick{\(|\zeta\rangle=\alpha_0|0\rangle+\alpha_1|1\rangle\)} & \ctrl{1} &\qw  & \gate{H} & \qw & \qw  \\
    \qw & \targ{} & \qw & \qw & \meter{} & \qw \\
     \lstick{\(|\nu\rangle=a_{00}|00\rangle+a_{11}|11\rangle\)} \Bigg{\{}\\
     \qw & \qw & \qw & \qw &\meter{} &  \qw   
\end{quantikz} \\
\caption{Teleportation circuit using usual non-maximally entangled
states}
\label{usualtele}
\end{figure}
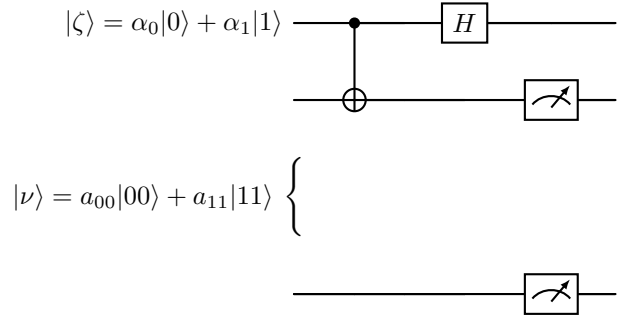

Let the usual information qubit be  \begin{equation}
 \label{info}
     |\zeta\rangle=\alpha_0|0\rangle+\alpha_1|1\rangle
 \end{equation}
 where $\alpha_0^2+\alpha_1^2=1$ . To send the information qubit via a  quantum teleportation channel we use the {\it usual non-maximally entangled state}  defined as :
 \begin{equation}
     \label{nu}
     |\nu\rangle=a_{00}|00\rangle+a_{11}|11\rangle
 \end{equation},
($a_{00}^2+a_{11}^2=1$).
where  $a_{00}\neq 1/\sqrt {2}$ , $a_{11}\neq 1/\sqrt{2}$ are the probability amplitudes and $DetA=a_{00}.a_{11}\neq 1/2$ . Consider the teleportation protocol Figure ( \ref{usualtele} ). 
Consider the teleportation protocol Figure (\ref{teledefbell}). The initial state that is being used as input in the circuit is $ |\zeta_0\rangle=|\zeta\rangle \otimes|\nu\rangle$. So we have :
            \begin{equation}
            \label{teleportation}
                \begin{split}
                    |\zeta_0\rangle=|\zeta\rangle \otimes|\nu\rangle=(\alpha_0|0\rangle +\alpha_1|1\rangle) \otimes (a_{00}|00\rangle+a_{11}|11\rangle)\\\xrightarrow[]{CNOT} [ \alpha_0|0\rangle \otimes (a_{00}|00\rangle+a_{11}|11\rangle) ]+[\alpha_1|1\rangle \otimes (a_{00}|10\rangle+a_{11}|01\rangle) ]
                  \\  \xrightarrow[]{HADAMARD} |\zeta_f\rangle=\frac{1}{\sqrt{2}}[|00\rangle(\alpha_0 a_{00}|0\rangle+\alpha_1 a_{11}|1\rangle) +|11\rangle(\alpha_0 a_{11}|1\rangle-\alpha_1 a_{00}|0\rangle) \\
                    +|01\rangle(\alpha_0 a_{11}|1\rangle+\alpha_1 a_{00}|0\rangle)
                    +|10\rangle(\alpha_0 a_{00}|0\rangle-\alpha_1 a_{11}|1\rangle)]
                \end{split}
            \end{equation}
          
          Now if alice (sender) measures the final state $|\zeta_f\rangle$ in basis $|00\rangle$, then Bob's (reciever) measurement will be, \begin{equation}
          \label{measure_susal}
            \begin{split}
                M_0=|\langle 0| \zeta_f\rangle|^2=\alpha_0^2 a_{00}^2/2 \enskip; \enskip M_1=|\langle 1 | \zeta_f\rangle|^2=\alpha_1^2 a_{11}^2/2 \\ and \enskip M_0 M_1=\alpha_0^2 \alpha_1^2 |DetA|^2/4
                \end{split}
            \end{equation}
			where $M_0$ and $M_1$ are Bob's (reciever) measurements in $|0\rangle$ and $|1\rangle$ basis respectively. $M_0M_1$ does not change even if Alice (sender) makes her measurement in any of the other three basis \{$|01\rangle,|10\rangle,|11\rangle$\}.
        The fidelity of the circuit is
            \begin{equation}
                \label{fidelity_susal}
                F= \langle \zeta_0|\rho_f|\zeta_0\rangle=(a_{00}\alpha_0+a_{11}\alpha_1)^2
                = (a_{00}\alpha_0 + \sqrt{1- a_{00}^2}\enskip \alpha_1)^2
            \end{equation}
            where $ \rho_f= |\zeta_f\rangle {}_f \langle \psi|$.
        For the fidelity to be an extremum;
              $  \frac{dF}{da_{00}}=0$ 
            i.e. $a_{00}=\pm \alpha_0 ;\pm \alpha_1$, so the two extremal fidelities ($F_{min};F_{max}$) are 
            \begin{equation}
            \label{F1&F2}
                F_{min}=4|DetA|^2 \enskip for \enskip a_{00}=\pm \alpha_1 \enskip \text{and} \enskip F_{max}=1 \enskip for \enskip a_{00}=\pm \alpha_0
          \end{equation}
          The second order derivative of F i.e $\frac{d^2F}{da_{00}^2}=2\frac{(2|\alpha_0|^2-1)^2}{|\alpha_0|^2}>0$ for $a_{00}=\pm \alpha_1$ and $\frac{d^2F}{da_{00}^2}=\frac{-2}{|\alpha_1|^2} <0$ when $a_{00}=\pm \alpha_0$, since $|\alpha_0|^2$ and $|\alpha_1|^2$ is always positive. So $F_{min}$ is the minimum fidelity and $F_{max}$ is the maximum fidelity. Thus we show that teleportation using non-maximally entangled state is possible in our formalism. In the usual case Alice (sender) has to tell Bob (receiver) through classical communication channel in which of the basis \{$| 00\rangle,|01\rangle,|10\rangle,|11\rangle$\}  she has made her own measurement and accordingly Bob (receiver) makes his measurements. In our protocol additionally $|DetA|$ must also be given to Bob (receiver) along with the basis in which Alice (sender) has made her measurement. Then using  $\alpha_0^2+\alpha_1^2=1$ and equation (\ref{measure_susal}) Bob (receiver) can calculate $\alpha_0$ and $\alpha_1$. More over the fidelity lies between $F_{min}$and $F_{max}$. For the maximally entangled state $|DetA|^2=1/4$ and  we get F=1 (which is an already known result).
  
 \section{Teleportation using non-maximally entangled q-deformed states:}
 {\it Case 1 : Undeformed information qubit $|\zeta\rangle$ using deformed entangled state $|\nu\rangle_q$ \\( equation \ref{muq})} \\

 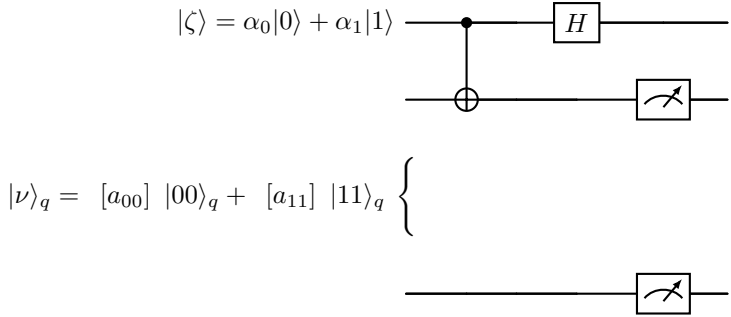
\begin{figure}
\begin{quantikz} 
    \lstick{\(|\zeta\rangle=\alpha_0|0\rangle+\alpha_1|1\rangle\)} & \ctrl{1} &\qw  & \gate{H} & \qw & \qw  \\
    \qw & \targ{} & \qw & \qw & \meter{} & \qw \\
     \lstick{\(|\nu\rangle_q=\enskip[a_{00}]\enskip|00\rangle_q+\enskip[a_{11}]\enskip|11\rangle_q\)} \Bigg{\{}\\
     \qw & \qw & \qw & \qw &\meter{} &  \qw   
\end{quantikz} \\
\caption{Teleportation Circuit using deformed entangled state}
\label{teledefbell}
\end{figure}

Consider a {\it non-maximally entangled bipartite state}  in Hilbert space $\mathcal{H}_{AB}$ .The operators  $d_i$ and $h_j$ are usual annihilation operators for the first and second oscillators  respectively and $d_{iq}$ and $h_{jq} $ are their deformed counterparts. The  same holds for the  creation operators and  $i,j=1,2$. Note these sets of operators define $|\nu\rangle_q$ .From equation (\ref{q-realization}) :    
\begin{equation}
\label{d,h}
    \begin{split}
    d_{1q}^{\dagger}=\sqrt{\bigg(\frac{q^{T_1}\omega_1(q)-q^{-T_1}\omega_2(q)}{{T_1(q-q^{-1})}}\bigg)}d_1^{\dagger} \enskip;\enskip 
		d_{2q}^{\dagger}=\sqrt{\bigg(\frac{q^{T_2}\omega_3(q)-q^{-T_2}\omega_4(q)}{{T_2(q-q^{-1})}}\bigg)}d_2^{\dagger} \\ h_{1q}^{\dagger}=\sqrt{\bigg(\frac{q^{R_1}\delta_1(q)-q^{-R_1}\delta_2(q)}{{R_1(q-q^{-1})}}\bigg)}h_1^{\dagger} \enskip ; \enskip h_{2q}^{\dagger}=\sqrt{\bigg(\frac{q^{R_2}\delta_3(q)-q^{-R_2}\delta_4(q)}{{R_2(q-q^{-1})}}\bigg)}h_2^{\dagger} 
        \end{split}
			\end{equation}
			 with $ T_{1q}=T_1-(1/s)ln(\omega_2(q))$ , $ T_{2q}=T_2-(1/s)ln(\omega_4(q))$ , $ R_{1q}=R_1-(1/s)ln(\delta_2(q))$, 
			 $ R_{2q}=R_2-(1/s)ln(\delta_4(q))$. Here $T_i,R_j$ (for i,j=1,2) are number operators of the usual harmonic oscillators and $T_{iq} , R_{jq}$ are the deformed ones. As $\omega(q)$ and $\delta(q)$ are arbitrary we take
$\omega_1(q)=\omega_2(q)=\omega_3(q)=\omega_4(q)=\omega(q)$ and $\delta_1(q)=\delta_2(q)=\delta_3(q)=\delta_4(q)=\delta(q)$. So equation (\ref{d,h}) becomes 
            \begin{equation}
            \label{d's,h's}
            \begin{split}
d_{1q}^{\dagger}=\sqrt{\bigg(\frac{q^{T_1}\omega(q)-q^{-T_1}\omega(q)}{{T_1(q-q^{-1})}}\bigg)}d_1^{\dagger} \enskip ;\enskip
d_{2q}^{\dagger}=\sqrt{\bigg(\frac{q^{T_2}\omega(q)-q^{-T_2}\omega(q)}{{T_2(q-q^{-1})}}\bigg)}d_2^{\dagger}\\
h_{1q}^{\dagger}=\sqrt{\bigg(\frac{q^{R_2}\delta(q)-q^{-R_2}\delta(q)}{{R_2(q-q^{-1})}}\bigg)}h_1^{\dagger} \enskip;\enskip 
h_{2q}^{\dagger}= \sqrt{\bigg(\frac{q^{R_2}\delta(q)-q^{-R_2}\delta(q)}{{R_1(q-q^{-1})}}\bigg)}h_2^{\dagger}
                \end{split}
			\end{equation}
$ T_{1q}=T_1-(1/s)ln(\omega(q))$ ,
$ T_{2q}=T_2-(1/s)ln(\omega(q))$ ,
$ R_{1q}=R_1-(1/s)ln(\delta(q))$ 
and $ R_{2q}=R_2-(1/s)ln(\delta(q))$ . \\
We define a non-maximally entangled q-deformed bipartite state as deformation of equation (\ref{nu} ) 
\begin{equation}
\label{muq}
\begin{split}
|\nu\rangle_q=\enskip[a_{00}]\enskip|00\rangle_q+\enskip[a_{11}]\enskip|11\rangle_q
\end{split}
\end{equation}
 where $\enskip [a_{00}]\enskip,[a_{11}]\enskip$ are the deformed amplitudes associated with the deformed basis $|00\rangle_q ,|11\rangle_q$ respectively and equation (\ref{d's,h's}) are the relevant creation operators. For the state  $|\nu\rangle_q$ to be non-maximally entangled we have  $\enskip[a_{00}]\enskip\neq \enskip[\frac{1}{\sqrt{2}}]\enskip$ , $\enskip[a_{11}]\enskip\neq \enskip[\frac{1}{\sqrt{2}}]\enskip$ and $DetA_q=\enskip[a_{00}]\enskip\enskip[a_{11}]\enskip\neq 0$. Also for normalization (equation \ref{psibeta}) $\omega(q)\delta(q)=\frac{1}{\enskip[a_{00}]^2+\enskip[a_{11}]^2}$.
Using equations (\ref{qubitoscillator}) and (\ref{d's,h's})  along with the fact that  the vacuum state is same for both usual and deformed oscillators we get:
 \begin{equation}
 \label{nuq}
\begin{split}
|\nu\rangle_q=\enskip[a_{00}]\enskip|00\rangle_q+\enskip[a_{11}]\enskip|11\rangle_q\\=\sqrt{\omega(q)\delta(q)} \Bigg(\enskip[a_{00}]\enskip|00\rangle+\enskip[a_{11}]\enskip|11\rangle \Bigg)
\end{split}
\end{equation}
 Note that   $ \lim_{q \to 1}|\nu\rangle_q =|\nu\rangle$.

In this section we will develop a protocol to enhance cryptographic security by sending the { \it undeformed} information qubit  using non-maximally entangled q-deformed states. We will take into account the fact that measurement can be done only in the usual basis. 
 In this  protocol Alice (sender) sends a single information qubit $|\zeta\rangle=\alpha_0|0\rangle+\alpha_1|1\rangle$ , where ($\alpha_0^2+\alpha_1^2=1$), to Bob (receiver) via quantum teleportation channel with a {\it q-deformed non-maximally entangled state $|\nu\rangle_q$ }  defined in equation (\ref{nuq}).

 Alice (sender) can make her measurement in any one  of the basis states $(|00\rangle,|01\rangle,|10\rangle,|11\rangle$). 

Consider the teleportation protocol Figure (\ref{teledefbell}). The initial state that is being used as input in the circuit is $ |\zeta_0\rangle_q=|\zeta\rangle \otimes|\nu\rangle_q$. So we have :
            \begin{equation}
            \label{teleportation}
            \begin{split}
|\zeta_0\rangle_q=|\zeta\rangle\otimes|\nu\rangle_q=\sqrt{\omega(q)\delta(q)}\Bigg((\alpha_0|0\rangle +\alpha_1|1\rangle) \otimes ([a_{00}]|00\rangle+[a_{11}]|11\rangle)\Bigg)  \\\xrightarrow[]{CNOT} \sqrt{\omega(q)\delta(q)}\bigg( \alpha_0|0\rangle \otimes ([a_{00}]|00\rangle+[a_{11}]|11\rangle) +\alpha_1|1\rangle \otimes ([a_{00}]|10\rangle+[a_{11}]|01\rangle) \bigg) 
                  \\  \xrightarrow[]{H} |\zeta_f\rangle_{q}=\frac{\sqrt{\omega(q)\delta(q)}}{\sqrt{2}}\bigg(|00\rangle(\alpha_0 [a_{00}]|0\rangle+\alpha_1 [a_{11}]|1\rangle) +|11\rangle(\alpha_0 [a_{11}]|1\rangle-\alpha_1 [a_{00}]|0\rangle)  \\
                    +|01\rangle(\alpha_0 [a_{11}]|1\rangle+\alpha_1[a_{00}]|0\rangle)
                    +|10\rangle([\alpha_0][ a_{00}]|0\rangle-\alpha_1 [a_{11}]|1\rangle)\bigg)
                    \end{split}
            \end{equation}
 Now if we take an example where Alice (sender) measures the final state $|\zeta_f\rangle_q$ in basis $|00\rangle$, then Bob (receiver)'s measurement will be, \begin{equation}
          \label{umeasure}
            \begin{split}
                M_0=|\langle 0| \zeta_f\rangle_q|^2=\omega(q)\delta(q)\alpha_0^2 \enskip[a_{00}]^2/2 \\ M_1=|\langle 1 | \zeta_f\rangle_q|^2=\omega(q)\delta(q)\alpha_1^2 \enskip[a_{11}]^2/2 \\  \enskip M_0 M_1=\omega^2(q)\delta^2(q)\alpha_0^2 \alpha_1^2 |DetA_q|^2/4
                \end{split}
            \end{equation}
			where $M_0$ and $M_1$ are Bob (receiver)'s measurements in $|0\rangle$ and $|1\rangle$ basis respectively. Also note that $M_0M_1$ does not change even if Alice (sender) makes her measurement in any of the other three basis ($|01\rangle,|10\rangle,|11\rangle$)

            In the usual case Alice (sender) has to tell Bob (receiver) through classical communication channel in which of the basis ($|
 00\rangle,|01\rangle,|10\rangle,|11\rangle$)  she has made her own measurement and accordingly Bob (receiver) makes his measurements. In our protocol additionally $|DetA_q|$ , exact form of $\omega(q),\delta(q)$ and $q=e^s , 
 (0\leq s\leq1)$ must also be given to Bob (receiver) along with the basis in which Alice (sender) has made her measurement. Then using $\alpha_0^2+\alpha_1^2=1$ and equation (\ref{umeasure}) Bob (receiver) can calculate $\alpha_0$ and $\alpha_1$.
           
            The fidelity of the circuit is
            \begin{equation}
             \label{fidelity1}
                F= {}_q\langle \zeta_0|\rho_{fq}|\zeta_0\rangle_q=\omega(q)\delta(q)(\enskip[a_{00}]\enskip\alpha_0+\enskip[a_{11}]\enskip\alpha_1)^2
                            \end{equation}
            where $ \rho_{fq}= |\zeta_f\rangle_q {}_q \langle \zeta_f|$.

Moreover for maximum fidelity {\it viz} $F =1$ at $q \to 1$  that is $\omega(q)\rightarrow1 $ and $\delta(q) \rightarrow 1$ the following must be satisfied.
\begin{equation}
\label{fid_max1}
    a_{00}\alpha_0+a_{11}\alpha_1=1
    \end{equation}
It is to be noted that the information is sent by realizing the entangled state  with q-deformed harmonic oscillators but once the information is received by Bob (receiver), he goes to the limit   $q \to 1$ to extract the original undeformed state which carries the information. So we get back the same condition for the maximum fidelity that we should get without any kind of q-deformation already expressed in equation (\ref{fidelity_susal}) and (\ref{fid_max1}). The bonus is that the teleportation has now become more secure from eavesdroppers.
\\

 {\it Case 2: Deformed information qubit $|\zeta\rangle_q$  using deformed entangled state  $|\nu\rangle_q$ } \\

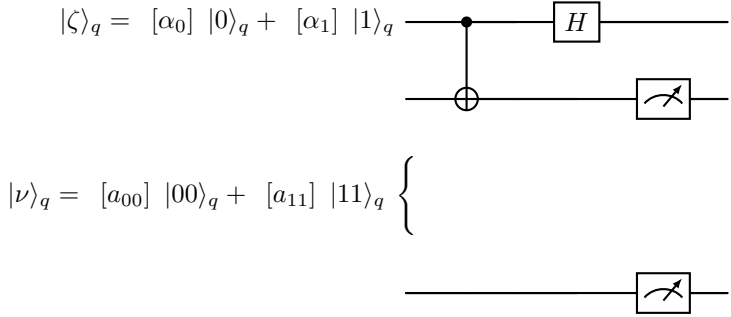
\begin{figure}
\begin{quantikz} 
    \lstick{\(|\zeta\rangle_q=\enskip[\alpha_0]\enskip|0\rangle_q+\enskip[\alpha_1]\enskip|1\rangle_q\)} & \ctrl{1} &\qw  & \gate{H} & \qw & \qw  \\
    \qw & \targ{} & \qw & \qw & \meter{} & \qw \\
     \lstick{\(|\nu\rangle_q=\enskip[a_{00}]\enskip|00\rangle_q+\enskip[a_{11}]\enskip|11\rangle_q\)} \Bigg{\{}\\
     \qw & \qw & \qw & \qw &\meter{} &  \qw   
\end{quantikz} \\
\caption{Teleportation Circuit using both deformed information and deformed entangled state}
\label{teledefqubit}
\end{figure}

If the creation and annihilation operators for the usual harmonic oscillator corresponding to the single information qubit $|\zeta\rangle$ be $c^{\dagger},c$ respectively and $c^{\dagger}_q ,c_q$ are the q-deformed creation and annihilation operators for the deformed information qubit $|\zeta\rangle_q$ respectively then according to above discussion (equation \ref{q-realization}) the relationship  between the deformed and original creation operators are as follows:
 \begin{equation}
    \begin{split}
    \label{infoq1}
    c_{1q}^{\dagger}=\sqrt{(\frac{q^{P_1}\gamma_1(q)-q^{-P_1}\gamma_2(q)}{{P_1(q-q^{-1})}})}c_1^{\dagger} \enskip;\enskip c_{1q}=c_1 \sqrt{(\frac{q^{P_1}\gamma_1(q)-q^{-P_1}\gamma_2(q)}{{P_1(q-q^{-1})}})}\\
		c_{2q}^{\dagger}=\sqrt{(\frac{q^{P_2}\gamma_3(q)-q^{-P_2}\gamma_4(q)}{{P_2(q-q^{-1})}})}c_2^{\dagger} \enskip;\enskip c_{2q}=c_2 \sqrt{(\frac{q^{P_2}\gamma_3(q)-q^{-P_2}\gamma_4(q)}{{P_2(q-q^{-1})}})}
        \end{split}
			\end{equation}
		where the number operators for the usual oscillator $P_i$ and that of deformed oscillator $P_{iq}$ (i=1,2) follows the relation $ P_{1q}=P_1-(1/s)ln(\gamma_2)$ and $ P_{2q}=P_2-(1/s)ln(\gamma_4)$ and $\gamma_k$'s (k=1,2,3,4) are arbitrary functions od $q$. If we choose $\gamma_i=\gamma$ (i=1,2,3,4) , then equation (\ref{infoq1}) becomes:
        \begin{equation}
    \begin{split}
    \label{infoq2}
    c_{1q}^{\dagger}=\sqrt{(\frac{q^{P_1}\gamma(q)-q^{-P_1}\gamma(q)}{{P_1(q-q^{-1})}})}c_1^{\dagger} \enskip;\enskip c_{1q}=c_1 \sqrt{(\frac{q^{P_1}\gamma(q)-q^{-P_1}\gamma(q)}{{P_1(q-q^{-1})}})}\\
		c_{2q}^{\dagger}=\sqrt{(\frac{q^{P_2}\gamma(q)-q^{-P_2}\gamma(q)}{{P_2(q-q^{-1})}})}c_2^{\dagger} \enskip;\enskip c_{2q}=c_2 \sqrt{(\frac{q^{P_2}\gamma(q)-q^{-P_2}\gamma(q)}{{P_2(q-q^{-1})}})}
        \end{split}
			\end{equation}
        So, the relation between the usual and deformed number operator becomes $ P_{1q}=P_1-(1/s)ln(\gamma)$ and $ P_{2q}=P_2-(1/s)ln(\gamma)$.

We now define the q-deformed information qubit as deformation of equation ( \ref{info} )
\begin{equation}
    \label{q-info}  |\zeta\rangle_q=\enskip[\alpha_0]\enskip|0\rangle_q+\enskip[\alpha_1]\enskip|1\rangle_q
\end{equation}
where $[\alpha_0]$ and $[\alpha_1]$ are the amplitudes associated with the deformed qubit $|0\rangle_q$ and $|1\rangle_q$ respectively and equation (\ref{infoq2}) are the relevant creation operators . (Note that $ \lim_{q \to 1}|\zeta\rangle_q =|\zeta\rangle$).
Using equations (\ref{qubitoscillator}) and (\ref{infoq2})  along with the fact that  the vacuum state is same for both usual and deformed oscillators we get:
\begin{equation}
   \label{q-info1}
\begin{split}
|\zeta\rangle_q=\enskip[\alpha_0]\enskip|0\rangle_q+\enskip[\alpha_1]\enskip|1\rangle_q \\
=\sqrt{\gamma(q)} \Bigg( \enskip[\alpha_0]\enskip|0\rangle+\enskip[\alpha_1]\enskip|1\rangle \Bigg)
    \end{split}
\end{equation}
In this section we will develop a protocol to enhance cryptographic security by sending the { \it q-deformed} information qubit  using {\it non-maximally entangled q-deformed states}. We will take into account the fact that measurement can be done only in the usual basis. 
        The q-deformed information qubit that is to be teleported using this protocol is  $|\zeta\rangle_q$ defined in equation (\ref{q-info1}) where numbers $\alpha_0$ and $\alpha_1$ has the original information that is to be sent. Two facts to be noted is that (a) for the normalization of the single deformation qubit $\gamma=\frac{1}{\enskip[\alpha_0]^2+\enskip[\alpha_1]^2}$ must be satisfied. (b) Under limiting condition $q\to 1$ the deformed information qubit $|\zeta\rangle_q \to |\zeta\rangle$ where $|\zeta\rangle$ is the original information qubit expressed in equation (\ref{info}).
 The {\it non-maximally entangled state expressed using q-deformation} that is being used in this protocol has already been defined in equation (\ref{nuq}).
  Alice (sender) can make her measurement in any one  of the basis states $(|00\rangle,|01\rangle,|10\rangle,|11\rangle$).
  \newpage
    
Consider the teleportation protocol Figure (\ref{teledefqubit}). The initial state that is being used as input in the circuit is $ |\zeta_0'\rangle_q=|\zeta\rangle_q \otimes|\nu\rangle_q$. So we have :
            \begin{equation}
            \begin{split}
                        \label{teleportation}
                    |\zeta_0'\rangle_q=|\zeta\rangle_q \otimes|\nu\rangle_q=\sqrt{\gamma(q)\omega(q)\delta(q)}\Bigg(([\alpha_0]|0\rangle +[\alpha_1]|1\rangle) \otimes ([a_{00}]|00\rangle+[a_{11}]|11\rangle) \Bigg)  \\\xrightarrow[]{CNOT} \sqrt{\gamma(q)\omega(q)\delta(q)} \bigg( [\alpha_0]|0\rangle \otimes ([a_{00}]|00\rangle+[a_{11}]|11\rangle) +[\alpha_1]|1\rangle   \otimes ([a_{00}]|10\rangle+[a_{11}]|01\rangle) \bigg)
                \\   \xrightarrow[]{H} |\zeta_f'\rangle_{q}=\frac{\sqrt{\gamma(q)\omega(q)\delta(q)}}{\sqrt{2}}\bigg(|00\rangle([\alpha_0] [a_{00}]|0\rangle+[\alpha_1][a_{11}]|1\rangle)  +|11\rangle([\alpha_0 ][a_{11}]|1\rangle\\-[\alpha_1][a_{00}]|0\rangle) 
                    +|01\rangle([\alpha_0]\ [a_{11}]|1\rangle+[\alpha_1][a_{00}]|0\rangle)
                   +|10\rangle([\alpha_0][ a_{00}]|0\rangle-[\alpha_1] [a_{11}]|1\rangle)\bigg)
                \end{split}
                        \end{equation}
 Now if Alice (sender) measures  the final state $|\zeta'_f\rangle_q$ in basis $|00\rangle$, then Bob (receiver)'s measurement will be, \begin{equation}
          \label{measure}
            \begin{split}
                M_0=|\langle 0| \zeta_f'\rangle_q|^2=\gamma(q)\omega(q)\delta(q)\enskip[\alpha_0]^2 \enskip[a_{00}]^2/2 \\ M_1=|\langle 1 | \zeta_f'\rangle_q|^2=\gamma(q)\omega(q)\delta(q)\enskip[\alpha_1]^2 \enskip[a_{11}]^2/2 \\ \enskip M_0 M_1=\gamma^2(q)\omega^2(q)\delta^2(q)\enskip[\alpha_0]^2 \enskip[\alpha_1]^2 |DetA_q|^2/4
                \end{split}
            \end{equation}
			where $M_0$ and $M_1$ are Bob (receiver)'s measurements in $|0\rangle$ and $|1\rangle$ basis respectively. $M_0M_1$ does not change even if Alice (sender) makes her measurement in any of the other three basis ($|01\rangle,|10\rangle,|11\rangle$)

            In the usual case Alice (sender) has to tell Bob (receiver) through classical communication channel in which of the basis ($| 00\rangle,|01\rangle,|10\rangle,|11\rangle$)  she has made her own measurement and accordingly Bob (receiver) makes his measurements. In our protocol additionally $|DetA_q|$ , exact form of $\omega(q),\delta(q),\gamma(q)$ , $q$ and $s$ must also be given to Bob (receiver) along with the basis in which Alice (sender) has made her measurement. Then using $\gamma(q)=\frac{1}{\enskip[\alpha_0]^2+\enskip[\alpha_1]^2}$ and equation (\ref{measure}) Bob (receiver) can calculate $\alpha_0$ and $\alpha_1$.Therefore our protocol has a large number of additional parameters which must be communicated to Bob (receiver). This enhances the cryptographic security of our protocol.

            The fidelity of the circuit is
            \begin{equation}
             \label{fidelity}
                F= {}_q\langle \zeta_0'|\rho_{fq}'|\zeta_0'\rangle_q=\gamma(q)\omega(q)\delta(q)(\enskip[a_{00}]\enskip\enskip[\alpha_0]\enskip+\enskip[a_{11}]\enskip\enskip[\alpha_1]\enskip)^2
                            \end{equation}
            where $ \rho_{fq}= |\zeta_f'\rangle_q {}_q \langle \zeta_f'|$.
It can be noted that fidelity of the teleportation circuit is a complicated function of $\omega(q),\gamma(q),\delta(q)$ and $\enskip[\alpha_0]\enskip,\enskip[\alpha_1]\enskip,\enskip[a_{00}]\enskip,\enskip[a_{11}]\enskip$ which are again function of $q$ where $q=e^s$.
Moreover for maximum fidelity {\it viz} $F =1$ at $q \to 1$ that is $\omega(q)\rightarrow1 $, $\delta(q) \rightarrow 1$ and $ \gamma(q)\rightarrow 1$ the following condition (\ref{fid_max}) must be satisfied.
\begin{equation}
\label{fid_max}
    a_{00}\alpha_0+a_{11}\alpha_1=1
    \end{equation}
It is to be noted that the information is sent by realizing the entangled state and the information qubit with q-deformed harmonic oscillators but once the information is received by Bob (receiver), he goes to the limit   $q \to 1$ to extract the original undeformed state which carries the information. So we get back the same condition for the maximum fidelity that we should get without any kind of q-deformation.The bonus is that the teleportation has now become more secure from eavesdroppers.

For the above circuit we take a very specific choice for the entangled state $|\nu\rangle_q=\enskip[a_{00}]\enskip\enskip|00\rangle_q+\enskip[a_{11}]\enskip\enskip|11\rangle_q$ but any other choice of entangled state like $|\nu'\rangle_q=\enskip[a_{01}]\enskip\enskip|01\rangle_q+\enskip[a_{10}]\enskip\enskip|10\rangle_q$, $|\nu''\rangle_q=\enskip[a_{01}]\enskip\enskip|01\rangle_q-\enskip[a_{10}]\enskip\enskip|10\rangle_q$,$|\nu'''\rangle_q=\enskip[a_{00}]\enskip\enskip|00\rangle_q-\enskip[a_{11}]\enskip\enskip|11\rangle_q$ can also be used.
\section{Conclusion:}
In this work we have employed the powerful techniques of quantum groups and q-deformed algebras to enhance the cryptographic security during teleportation. To achieve this we first constructed q-deformed Bell-like  states which reduce to the usual Bell states when the deformation parameter $q\rightarrow 1$. We then show that these q-deformed Bell-like states form  an orthonormal basis for q-deformed entangled bipartite states when certain arbitrary functions of $q$ satisfy a constraint. Next we discuss the  generalisation of the usual teleportation protocol  with non-maximally entangled states and use this generalisation to construct two new protocols using   q-deformed non-maximally entangled  states. These states have additional parameters and these have to be shared for decryption after teleportation. Consequently, the cryptographic security is improved.

\section{Acknowledgement}
One of the author(PDG) would like to thank Sister Nivedita University for providing research scholarship, under Student ID-2331207003001

\end{document}